%% file: l4dc2023-sample.tex
\documentclass[12pt, final]{l4dc2023}

\title[Equilibria in Decentralized Learning]{Equilibria of Fully Decentralized Learning in Networked Systems}
\usepackage{times}

\usepackage{amsmath}

\usepackage{amssymb}
\usepackage{amsfonts}
\usepackage{graphicx}
\usepackage{dsfont}
\usepackage{mathrsfs}
\usepackage{siunitx}
\usepackage{indentfirst}
\usepackage{threeparttable}
\usepackage{multirow}
\providecommand{\proofname}{Proof}
\newenvironment{Proof}%
{%
\par\noindent{\bfseries\upshape \proofname\ }%
}%
{\jmlrQED}
\newcommand{\real}[0]{\mathbb R}

\DeclareSymbolFont{bbold}{U}{bbold}{m}{n}
\DeclareSymbolFontAlphabet{\mathbbold}{bbold}

\newcommand{\diag}[1]{\ensuremath{\mathrm{diag}(#1)}}
\newcommand{\tr}[1]{\ensuremath{\mathrm{tr}\left(#1\right)}}

\usepackage{tikz}
\newcommand*\circled[1]{\tikz[baseline=(char.base)]{\node[shape=circle,draw,inner sep=0.05pt] (char) {#1};}}

\newtheorem{thm}{Theorem}

\newtheorem{rem}{Remark}
\newtheorem{lem}{Lemma}
\newtheorem{ass}{Assumption}
\newtheorem{prop}{Proposition}
\newtheorem{conj}{Conjecture}
\let\labelindent\relax
\usepackage{enumitem}
\setlist[itemize]{leftmargin=*}
\allowdisplaybreaks

\author{%
 \Name{Yan {Jiang}, Wenqi {Cui}, Baosen Zhang} \Email{\{jiangyan,wenqicui,zhangbao\}@uw.edu}\\
 \addr Department
of Electrical and Computer Engineering, University of Washington, Seattle, WA 98195, USA
 \AND
 \Name{Jorge Cort\'es} \Email{cortes@ucsd.edu}\\
 \addr Department of Mechanical and Aerospace Engineering, University of California, San Diego, CA 92093, USA%
}

\begin{document}

\maketitle

\begin{abstract}%
Existing settings of decentralized learning either require players to have full information or the system to have certain special structure that may be hard to check and hinder their applicability to practical systems. To overcome this, we identify a structure that is simple to check for linear dynamical system, where each player learns in a fully decentralized fashion to minimize its cost. We first establish the existence of pure strategy Nash equilibria in the resulting noncooperative game. We then conjecture that the Nash equilibrium is unique provided that the system satisfies an additional requirement on its structure. We also introduce a decentralized  mechanism based on projected gradient descent to have agents learn the Nash equilibrium. Simulations on a $5$-player game validate our results.
\end{abstract}


\section{Introduction}
\input{01-intro.tex}
\section{Problem Setup}\label{sec:model}
%
%
\input{02-model.tex}

\section{Pure Strategy Nash Equilibria}\label{sec:nash}

In this section, we study the existence and uniqueness of pure strategy Nash equilibria for the $n$-player noncooperative
game introduced in Section~\ref{sec:model}.
\vspace{-0.2cm}
\subsection{Existence of Nash Equilibrium}\label{sec:existence}
\input{03-nash.tex}

\section{Decentralized Learning of Nash Equilibrium via Projected Gradient Descent}\label{sec:learning}

\input{04-learning.tex}

\section{Conclusions and Outlook}\label{sec:conclusion}
We have formulated a fully decentralized learning problem for a symmetric linear dynamical system as a noncooperative game. We have shown the existence of pure strategy Nash equilibrium and conjectured its uniqueness under additional conditions on the state matrix. We have used projected gradient descent to have agents learn, in a fully decentralized way, the Nash equilibrium. Simulations in a $5$-player game confirm our conjecture on the uniqueness of Nash equilibrium. Future work will explore the uniqueness of Nash equilibria in the general case, the analysis of the noncooperative game when the distribution of the initial states is not white, the formal characterization of the robustness of the proposed action updating rule, and
the extension of the results to time-varying topologies. 



\acks{This material is based in part upon work supported by the State of Washington through the University of Washington Clean Energy Institute. Yan Jiang, Wenqi Cui, and Baosen Zhang were partially supported by NSF Award ECCS-2153937. Jorge Cort\'es was partially supported by  NSF Award IIS-2007141.}

\bibliography{Decentral}

\end{document}

%% file: 01-intro.tex
Many real world systems are too large and complex for decisions to be made in a centralized fashion. Instead, there are a multitude of decision makers (or players), interacting over a system, each possessing limited knowledge and observations. Thus, the study of decentralized decision making has been a topic of interest for several decades (see, e.g.,~\cite{tsitsiklis1984problems,olfati2007consensus}; and the references within). Recently, the advancement of machine learning in multi-agent setting has attracted significant attention from the control and learning communities, which gives rise to successful application of multi-agent decision learning across multiple fields, including power system operations~\citep{yang2018recurrent,cui2022decentralized}, traffic control~\citep{bazzan2009opportunities}, communication networks~\citep{han2012game}, and others~\citep{Li22}. 


A foundational question in decentralized decision making and learning is whether the players would reach some type of equilibria. One setting that has been extensively studied is that of linear quadratic (LQ) games~\citep{zhangNEURIPS2019, EricAAMAS2020}, which generalize the well-known linear quadratic regulator (LQR) problems. Unlike LQR problems, which can be considered to have a single agent~\citep{fazel2018global}, LQ games have multiple players that interact over a linear system and all try to minimize their individual regulation and control costs. This complexity makes LQ games not enjoy guarantees of convergence to Nash equilibria, a property enjoyed by the LQR problem with policy-gradient methods. Moreover, although a player is limited in its input channels~\citep{bacsar1998dynamic,engwerda2005lq,Li22}, it typically has full information (access to the full state). This assumption of the availability of full information makes it difficult to apply related results to many practical systems. 

For systems where the players are limited both in input and information, several properties have been discovered to guarantee that good decentralized controllers can be found in a tractable manner. These include quadratic invariance~\citep{rotkowitz2002decentralized,lessard2014algebraic}, spatial invariance~\citep{bamieh2002distributed}, partially nestedness~\citep{shah2013cal}, and positiveness~\citep{rantzer2015scalable}. These conditions, however, can be challenging to check in practice. In addition, they often require the players to have nested information, which may not hold in practice. 

It is important to note that some structure of the system is necessary for decentralized learning to be analytically tractable. Even for linear systems, finding the optimal controllers is  NP-hard in general~\citep{blondel2000survey}. Even if restricted to linear controllers, the feasible set of the stabilizing controllers can have an exponential number of connected pieces~\citep{feng2019exponential}, and understanding the equilibria or convergence of the learned controllers becomes very difficult. 

In this paper, we identify a new structure for a class of systems where the behavior of decentralized learning can be explicitly characterized. In particular, we study a game over a linear and symmetric dynamical system with each player being modeled as a part of the state, where the action of the players is to choose a linear feedback control gain that minimizes quadratic costs on its own state regulation and control effort. Notably, compared to existing LQ games in the literature~\citep{fazel2018global, zhangNEURIPS2019, EricAAMAS2020}, we adopt a fully decentralized setting, in the sense that each player only knows its own information and takes an action directly affecting its own state, which can be naturally characterized as a noncooperative game. This captures the fact that, in many settings, the edge devices are becoming increasingly intelligent and capable of making sophisticated decisions, but they still do not have access to regular real-time communication with other devices. We show that there exists at least one pure strategy Nash Equilibrium in such a noncooperative game by establishing that, in contrast to existing results, the stabilizing controllers lie in a convex region and the cost functions are convex in the control gains. This, in turn, provides a simple way for players to estimate the gradients of the cost functions in a completely decentralized way and converge to the Nash equilibrium through gradient play. We conjecture that the pure strategy Nash equilibria is in fact unique, where partial and numerical results for this conjecture are provided. 

The key property used in our analysis is the symmetry of the dynamical system. That is, the state matrix is symmetric. This condition is distinct from existing ones and has the benefit that it is simple to check. The symmetry of the system is satisfied by many practical systems. For example, in power distribution systems with angle droop control~\citep{zhang2016transient,huang2020holistic}, the system is always symmetric since the matrix comes from the Laplacian of the underlying network.

The rest of this paper is organized as follows. Section~\ref{sec:model} formulates the decentralized learning problem in a symmetric linear dynamical system as a $n$-player noncooperative game. Section~\ref{sec:nash} shows the existence of pure strategy Nash equilibria in such a game and further conjectures its uniqueness under additional conditions on the system structure. Section~\ref{sec:learning} introduces a decentralized learning mechanism based on projected gradient descent and discusses how to implement it. Section~\ref{sec:conclusion} presents our conclusion and ideas for future work.







%% file: 02-model.tex
We consider a networked system with $n$ players (or agents), whose 
dynamics is given by
\begin{equation}\label{eq:sys}
    \dot{\boldsymbol{x}}(t)= \boldsymbol{A}\boldsymbol{x}(t) + \boldsymbol{u}(t)\,,
\end{equation}
where $\boldsymbol{x}(t):=\left(x_i(t), i \in \left[n\right] \right) \in \real^n$ is the state vector, $\boldsymbol{u}(t):=\left(u_i(t), i \in \left[n\right] \right) \in \real^n$ is the control input, and $\boldsymbol{A}\in \real^{n \times n}$ is the state matrix.\footnote{Throughout the paper, vectors are denoted in bold lower case and matrices are denoted in bold upper case, while scalars are unbolded.} We make the following assumption:

\begin{ass}[Structure assumption]\label{ass:ass}
The state matrix $\boldsymbol{A}$ is symmetric and negative definite. 
\end{ass}
\begin{rem}[Structure interpretations and extensions]
{\rm The symmetry of the state matrix is key to our developments since most of results in this paper critically rely on this assumption. The negative definiteness can be relaxed as discussed later in Remark~\ref{rem:A-K}. Assumption~\ref{ass:ass} reflects the graph structure of the system found in many applications. For example, in microgrid control~\citep{huang2020holistic,Cui2022lossy}, $\boldsymbol{A}$ is related to a Laplacian matrix that captures the active power flow.}\hfill $\square$
\end{rem}

To model a fully decentralized setting, we assume that each controller $u_i$ can only depend on the state of the $i$th player. Namely,  each player chooses an action $k_i\in\left[0, \overline{k}_i\right]$, with $\overline{k}_i>0$ being some upper bound, such that the $i$th component of the control input is determined by
\begin{align}\label{eq:u}
u_i(t)=-k_ix_i(t)\,. 
\end{align}
Let $\boldsymbol{K}:=\diag{k_i,i \in \left[n\right]}\in\real^{n \times n}$. Then the closed-loop system of \eqref{eq:sys} under \eqref{eq:u} becomes 
\begin{equation}\label{eq:sys-cl}
\dot{\boldsymbol{x}}(t)= (\boldsymbol{A}-\boldsymbol{K})\boldsymbol{x}(t)\,. 
\end{equation}

\begin{rem}[Hurwitz closed-loop system matrix]\label{rem:A-K}
{\rm Note that, with Assumption~\ref{ass:ass} on $\boldsymbol{A}$ and the restrictions that $k_i\geq0$, the closed-loop system \eqref{eq:sys-cl} is always stable since $\boldsymbol{A}-\boldsymbol{K} \prec 0$. However, if $\boldsymbol{A}$ is symmetric but not negative definite, the set of controller gains that make the closed-loop system \eqref{eq:sys-cl} stable is a convex set determined by $\boldsymbol{A}-\boldsymbol{K} \prec 0$. In this case, it is easy to find a lower bound $\underline{k}_i$ such that $\boldsymbol{A}-\boldsymbol{K} \prec 0$ if $k_i>\underline{k}_i$, $\forall i \in \left[n\right]$. 
} \hfill $\square$
\end{rem}

The goal of the $i$th player is to minimize its own expected cost $J_i(k_i,k_{-i})$ on state deviations and control effort along the trajectories of the system \eqref{eq:sys-cl}, given the actions of other players  $k_{-i}:=\{k_1,\dots,k_{i-1},k_{i+1},\dots,k_n\}$. Formally, we define the cost of the $i$th player as
\begin{align}\label{eq:cost-def}
J_i(k_i,k_{-i}):=&\ \mathbb{E}\! \left[\int_{0}^\infty\!\!\left(x_i^2(t)+\rho_i u_i^2(t)\right)\!\!\ \mathrm{d}t\right] \,,\quad\forall i \in \left[n\right],
\end{align}
where $\rho_i\geq0$ is the coefficient for tradeoff between the two components (state deviation and control effort). Note that the expectation $\mathbb{E} \left[\cdot\right]$ is taken with respect to random initial conditions $\boldsymbol{x}(0)$, where we make the common assumption (e.g., see~\cite{mendel1971bibliography}) that $\mathbb{E} \left[\boldsymbol{x}(0) \boldsymbol{x}(0)^T\right]=\boldsymbol{I}_n$, that is, the components of $\boldsymbol{x}(0)$ are independent and identically distributed (i.i.d.). 
\begin{rem}[Infinite time-horizon]
{\rm We take an infinite time-horizon in \eqref{eq:cost-def} for several reasons. 
First, it makes the analysis cleaner and thus is adopted in many settings (e.g., see~\cite{fazel2018global,dean2020sample,bu2019lqr}; and references within). Second, for stable systems, finite trajectory costs are well approximated by an infinite trajectory cost if the number of time steps in a trajectory is not too small. Third, this cost is equivalent to the expected average cost in systems with persistent white noise~\citep{mendel1971bibliography,kwon2006receding,weitenberg2018robust}.}\hfill $\square$ 
\end{rem}

The setting above defines a noncooperative game, where each player has action space $k_i \in [0,\overline{k}_i]$ and cost $J_i(k_i,k_{-i})$. A pure strategy Nash equilibrium of the game is defined as an action profile of the players where no single player $i$ can obtain a lower cost by choosing a different action, given that the actions of other players are fixed. That is, $(k_1^*,\dots,k_n^*)$ is a pure strategy Nash equilibrium if, $\forall i \in \left[n\right]$, $J_i(k_i^*,k_{-i}^*) \leq J_i(k_i',k_{-i}^*)$, $\forall k_i' \in [0,\overline{k}_i]$.
It is well known that not all games have a pure strategy Nash equilibrium~\citep{bacsar1998dynamic}, especially when the actions of the players are not explicitly reflected in the cost functions~\citep{marden2015game}.
%
%
We study in Section~\ref{sec:nash} the existence of pure strategy Nash equilibria for this game and describe in Section~\ref{sec:learning} how the players update their actions to find them.

%% file: 03-nash.tex
Notice that each player has an action space $\left[0, \overline{k}_i\right]$ that is a closed, bounded, and convex subset of $\real$. Therefore, based on the well-known result~\citep[Theorem 4.3]{bacsar1998dynamic}, in order to show the existence of pure strategy Nash equilibria, it suffices to show that the cost function $J_i(k_i,k_{-i})$ is jointly continuous in all its arguments and strictly convex in $k_i$, for every $k_{-i}$. In this subsection, we proceed by presenting a sequence of results that eventually enable us to prove the following main result.
\begin{thm}[Existence of pure strategy Nash equilibria] \label{thm:existence-NE}
The $n$-player noncooperative game admits a pure strategy Nash equilibrium.
\end{thm}



We start by investigating an explicit expression for the cost function $J_i(k_i,k_{-i})$. Clearly, $k_i$ does not explicitly show up in the definition of $J_i(k_i,k_{-i})$ given in~\eqref{eq:cost-def}, which hinders our analysis. The next result addresses this by providing an explicit expression of $J_i(k_i,k_{-i})$ in terms of $k_i$.

\begin{lem}[Individual cost functions]\label{lem:Ji-ki}
The cost function of the $i$th player, $\forall i \in \left[n\right]$, is given by
\begin{align}\label{eq:Ji}
    J_i(k_i,k_{-i})=\dfrac{\left(1+\rho_i k_i^2\right)}{2} f_i(k_i,k_{-i}) \,, \quad\text{with } f_i(k_i,k_{-i}) := \boldsymbol{e}_i^T \left(\boldsymbol{K}-\boldsymbol{A}\right)^{-1}\boldsymbol{e}_i\,,
\end{align}
where $\boldsymbol{e}_i\in \real^n$ is the $i$th standard basis vector. 
\end{lem}
\begin{Proof}
First, substituting \eqref{eq:u} to \eqref{eq:cost-def} yields
\begin{align}\label{eq:Ji-xt}
    J_i(k_i,k_{-i})=\ \left(1+\rho_i k_i^2\right)\mathbb{E}\!\left[\int_{0}^\infty\!\! x_i^2(t)\ \mathrm{d}t\right]
    =\left(1+\rho_i k_i^2\right)\mathbb{E}\!\left[\int_{0}^\infty\!\! \boldsymbol{x}(t)^T \boldsymbol{e}_i \boldsymbol{e}_i^T\boldsymbol{x}(t)\ \mathrm{d}t\right]\,,
\end{align}
where the second equality uses $x_i(t) = \boldsymbol{e}_i^T \boldsymbol{x}(t)$. Note that $\boldsymbol{x}(t)$ in \eqref{eq:Ji-xt} still implicitly depends on $k_i$. Therefore, we need to get an explicit expression of $\boldsymbol{x}(t)$ in terms of $k_i$ to perform further analysis.

Since the solution to the closed-loop system \eqref{eq:sys-cl} is $\boldsymbol{x}(t)=e^{\left(\boldsymbol{A}-\boldsymbol{K}\right)t}\boldsymbol{x}(0)$, \eqref{eq:Ji-xt} becomes:
\begin{align}
    J_i(k_i,k_{-i})
    \stackrel{\circled{1}}{=}&\left(1+\rho_i k_i^2\right)\mathbb{E}\! \left[\int_{0}^\infty\!\! \boldsymbol{x}(0)^Te^{\left(\boldsymbol{A}-\boldsymbol{K}\right)t} \boldsymbol{e}_i \boldsymbol{e}_i^T e^{\left(\boldsymbol{A}-\boldsymbol{K}\right)t}\boldsymbol{x}(0)\ \mathrm{d}t\right]\nonumber\\
    =&\left(1+\rho_i k_i^2\right)\mathbb{E}\! \left[\boldsymbol{x}(0)^T\!\!\int_{0}^\infty\!\! e^{\left(\boldsymbol{A}-\boldsymbol{K}\right)t} \boldsymbol{e}_i \boldsymbol{e}_i^T e^{\left(\boldsymbol{A}-\boldsymbol{K}\right)t}\ \mathrm{d}t\ \boldsymbol{x}(0)\right]\nonumber\\
    =&\left(1+\rho_i k_i^2\right)\mathbb{E}\! \left[\tr{\boldsymbol{x}(0)^T\!\! \int_{0}^\infty\!\! e^{\left(\boldsymbol{A}-\boldsymbol{K}\right)t} \boldsymbol{e}_i \boldsymbol{e}_i^T e^{\left(\boldsymbol{A}-\boldsymbol{K}\right)t}\ \mathrm{d}t \ \boldsymbol{x}(0)}\right]\nonumber\\
    \stackrel{\circled{2}}{=}&\left(1+\rho_i k_i^2\right)\mathbb{E}\! \left[\tr{\int_{0}^\infty\!\! e^{\left(\boldsymbol{A}-\boldsymbol{K}\right)t} \boldsymbol{e}_i \boldsymbol{e}_i^T e^{\left(\boldsymbol{A}-\boldsymbol{K}\right)t}\ \mathrm{d}t\ \boldsymbol{x}(0)\boldsymbol{x}(0)^T}\right]\nonumber\\
    =&\left(1+\rho_i k_i^2\right)\tr{\int_{0}^\infty\!\! e^{\left(\boldsymbol{A}-\boldsymbol{K}\right)t} \boldsymbol{e}_i \boldsymbol{e}_i^T e^{\left(\boldsymbol{A}-\boldsymbol{K}\right)t}\ \mathrm{d}t\ \mathbb{E}\! \left[\boldsymbol{x}(0)\boldsymbol{x}(0)^T\right]}\nonumber\\
    \stackrel{\circled{3}}{=}&\left(1+\rho_i k_i^2\right)\tr{\int_{0}^\infty\!\! e^{\left(\boldsymbol{A}-\boldsymbol{K}\right)t} \boldsymbol{e}_i \boldsymbol{e}_i^T e^{\left(\boldsymbol{A}-\boldsymbol{K}\right)t}\ \mathrm{d}t}\nonumber\\
    \stackrel{\circled{4}}{=}&\left(1+\rho_i k_i^2\right)\tr{\int_{0}^\infty\!\!  \boldsymbol{e}_i^T e^{2\left(\boldsymbol{A}-\boldsymbol{K}\right)t}\boldsymbol{e}_i\ \mathrm{d}t}\nonumber\\
    =&\left(1+\rho_i k_i^2\right) \tr{ \boldsymbol{e}_i^T \int_{0}^\infty\!\! e^{2\left(\boldsymbol{A}-\boldsymbol{K}\right)t}\ \mathrm{d}t\boldsymbol{e}_i} =\left(1+\rho_i k_i^2\right) \boldsymbol{e}_i^T \int_{0}^\infty\!\! e^{2\left(\boldsymbol{A}-\boldsymbol{K}\right)t}\ \mathrm{d}t\ \boldsymbol{e}_i\,,\label{eq:cost-i}
\end{align}
where \circled{1} uses the closed-loop solution, \circled{2} and \circled{4} use the cyclic property of the trace, and \circled{3} uses $\mathbb{E} \left[\boldsymbol{x}(0) \boldsymbol{x}(0)^T\right] = \boldsymbol{I}_n$. Note that the integral $\int_{0}^\infty\!\! e^{2\left(\boldsymbol{A}-\boldsymbol{K}\right)t}\ \mathrm{d}t$ is a common integral (see, e.g., \cite{hespanha2018linear}) given by 
\begin{align}
\!\!\!\!\!\!\int_{0}^\infty\!\! e^{2\left(\boldsymbol{A}-\boldsymbol{K}\right)t}\ \mathrm{d}t
  =\dfrac{\left(\boldsymbol{K}\!-\!\boldsymbol{A}\right)^{-1}}{2}\,.\label{eq:int-matrix-exp}
\end{align}
Therefore, substituting \eqref{eq:int-matrix-exp} into \eqref{eq:cost-i} yields \eqref{eq:Ji}.
\end{Proof}

Lemma~\ref{lem:Ji-ki} provides an explicit expression of the cost function $J_i(k_i,k_{-i})$ in terms of the action~$k_i$. This allows us next to characterize the convex properties of the cost function.

\begin{lem}[Strict convexity of cost functions]\label{lem:convex}
The cost function $J_i(k_i,k_{-i})$, $\forall i \in \left[n\right]$, is strictly convex in $k_i$ for each $k_{-i}\in\prod_{j\in\left[n\right]\setminus\{i\}}\left[0, \overline{k}_j\right]$.
\end{lem}
\begin{Proof}
Note that the action space $\left[0, \overline{k}_i\right]$ of each player is convex. By the second-order condition for convexity~\citep[Chapter 3.1.4]{Boyd2004convex}, in order to establish the strict convexity of $J_i(k_i,k_{-i})$ in $k_i$, it suffices to show that
\begin{align}\label{eq:2nd-con}
\dfrac{\partial^2 J_i(k_i,k_{-i})}{\partial k_i^2}>0 \,,\qquad    \forall i \in \left[n\right] .
\end{align}
We start by taking the partial derivative of $J_i(k_i,k_{-i})$ with respect to $k_i$. Direct calculations on \eqref{eq:Ji} show that
\begin{align}\label{eq:dJi}
    \dfrac{\partial J_i(k_i,k_{-i})}{\partial k_i}=&\ \rho_i k_i f_i(k_i,k_{-i})+\dfrac{\left(1+\rho_i k_i^2\right)}{2}\dfrac{\partial f_i(k_i,k_{-i})}{\partial k_i}\,.
\end{align}
Next, we take the partial derivative of \eqref{eq:dJi} with respect to $k_i$ to obtain
\begin{align}\label{eq:ddJi}
    &\ \dfrac{\partial^2 J_i(k_i,k_{-i})}{\partial k_i^2}
    =\ \rho_i f_i(k_i,k_{-i})+2\rho_i k_i \dfrac{\partial f_i(k_i,k_{-i})}{\partial k_i}+\dfrac{\left(1+\rho_i k_i^2\right)}{2}\dfrac{\partial^2 f_i(k_i,k_{-i})}{\partial k_i^2}\,.
\end{align}
To obtain the partial derivative of $f_i(k_i,k_{-i})$, we use~\eqref{eq:Ji} to get
\begin{align}
    \dfrac{\partial f_i(k_i,k_{-i})}{\partial k_i}=&\ \boldsymbol{e}_i^T \dfrac{\partial\left[ \left(\boldsymbol{K}-\boldsymbol{A}\right)^{-1}\right]}{\partial k_i}\boldsymbol{e}_i\nonumber
    \stackrel{\circled{1}}{=}-\boldsymbol{e}_i^T\left(\boldsymbol{K}-\boldsymbol{A}\right)^{-1} \dfrac{\partial \left(\boldsymbol{K}-\boldsymbol{A}\right)}{\partial k_i}\left(\boldsymbol{K}-\boldsymbol{A}\right)^{-1}\boldsymbol{e}_i\nonumber\\=&-\boldsymbol{e}_i^T\left(\boldsymbol{K}-\boldsymbol{A}\right)^{-1} \boldsymbol{e}_i\boldsymbol{e}_i^T\left(\boldsymbol{K}-\boldsymbol{A}\right)^{-1}\boldsymbol{e}_i 
    \stackrel{\circled{2}}{=}-f^2_i(k_i,k_{-i})\,,\label{eq:dfi}
\end{align}
where \circled{1} uses the formula for the derivative of an inverse matrix~\citep[Chapter 2.2]{Petersen2012matrix} and \circled{2} uses the definition of $f_i(k_i,k_{-i})$ in \eqref{eq:Ji} twice. Further taking the partial derivative of \eqref{eq:dfi} with respect to $k_i$ yields
\begin{align}\label{eq:ddfi}
    \dfrac{\partial^2 f_i(k_i,k_{-i})}{\partial k_i^2}=-2f_i(k_i,k_{-i})\dfrac{\partial f_i(k_i,k_{-i})}{\partial k_i}=2f_i^3(k_i,k_{-i})\,,
\end{align}
where the second equality uses \eqref{eq:dfi}.
Now, substituting \eqref{eq:dfi} and \eqref{eq:ddfi} into \eqref{eq:ddJi} yields 
\begin{align}
    \dfrac{\partial^2 J_i(k_i,k_{-i})}{\partial k_i^2}
    =&\ \rho_i f_i(k_i,k_{-i})-2\rho_i k_i f^2_i(k_i,k_{-i})+\left(1+\rho_i k_i^2\right)f^3_i(k_i,k_{-i})\nonumber\\=&\ f_i(k_i,k_{-i})\left[\rho_i\left(1-k_i f_i(k_i,k_{-i})\right)^2+f^2_i(k_i,k_{-i})\right]\,.\label{eq:J2d}
\end{align}
Clearly, the sign of \eqref{eq:J2d} only depends on the sign of $f_i(k_i,k_{-i})$, since $\rho_i\geq0$ and the terms inside the square brackets are squared.
It follows from Remark~\ref{rem:A-K} that $\left(\boldsymbol{K}-\boldsymbol{A}\right)^{-1}\succ0$, which further implies that $f_i(k_i,k_{-i})>0$ by its definition in~\eqref{eq:Ji}. Hence, it follows directly from~\eqref{eq:J2d} that \eqref{eq:2nd-con} holds, concluding the proof of strict convexity.
\end{Proof}

We now have the core element required to establish the existence of pure strategy Nash equilibria stated in Theorem~\ref{thm:existence-NE}.

\begin{Proof}{\bf of Theorem~\ref{thm:existence-NE}.}
Recall that, by~\citep[Theorem 4.3]{bacsar1998dynamic}, since the action space $\left[0, \overline{k}_i\right]$ of each player is a closed, bounded, and convex subset of $\real$, the $n$-player noncooperative game in question admits a pure strategy Nash equilibrium if the cost function $J_i(k_i,k_{-i})$ is jointly continuous in all its arguments and strictly convex in $k_i$ for every $k_{-i}$. By Lemma~\ref{lem:convex}, $J_i(k_i,k_{-i})$ is strictly convex in $k_i$ for every $k_{-i}$. Thus, it remains to show that $J_i(k_i,k_{-i})$ is jointly continuous in all its arguments. This can be seen clearly when one notices that $J_i(k_i,k_{-i})$ in \eqref{eq:Ji} is eventually a quotient of two multivariable polynomial functions in $k_j$, $\forall j\in\left[n\right]$. First, since the multivariable polynomial functions can be considered as a sum of products of polynomial functions, their joint continuity directly follows from the fact that polynomial functions are continuous everywhere and a product or sum of continuous functions is continuous as well~\citep[Theorem 4.9]{Rudin2013PMA}. Then, a quotient of such two continuous functions is also continuous everywhere except perhaps at the points which make the denominator zero~\citep[Theorem 4.9]{Rudin2013PMA}. Yet, no such points exist in the action space since $\boldsymbol{K}-\boldsymbol{A}\succ0$ by Remark~\ref{rem:A-K}. Hence, $J_i(k_i,k_{-i})$ is jointly continuous in all its arguments, concluding the proof.
\end{Proof}

\subsection{Uniqueness of Nash Equilibrium}
Having shown the existence of pure strategy Nash Equilibria in the $n$-player noncooperative game, here we analyze its uniqueness. A well-established condition to guarantee uniqueness is proposed by Rosen~\citep[Theorem 6]{rosen1965Eco} and states that the game in question admits a unique Nash equilibrium if, $\forall\boldsymbol{k}:= \left(k_i, i \in \left[n\right] \right) \in\prod_{i\in\left[n\right]}\left[0, \overline{k}_i\right]$,
\begin{align}\label{eq:rosen}
\boldsymbol{G}(\boldsymbol{k})+\boldsymbol{G}^T(\boldsymbol{k})\succ0\,,
\end{align}
where $\boldsymbol{G}(\boldsymbol{k})\in \real^{n \times n}$ is the Jacobian of the so-called pseudogradient defined as the stacked vector of partial derivatives of the cost function $J_i(k_i,k_{-i})$ with respect to the action $k_i$, i.e.,
\begin{align}\label{eq:g}
\boldsymbol{g}(\boldsymbol{k}):= \left(\dfrac{\partial J_i(k_i,k_{-i})}{\partial k_i}, i \in \left[n\right] \right) \in \real^n\qquad\mbox{and}\qquad G_{ij}(\boldsymbol{k}):=\frac{\partial g_i (\boldsymbol{k})}{\partial k_j} \;. 
\end{align}
In general, the condition \eqref{eq:rosen} does not hold for the game in question. Yet, after performing extensive numerical tests, we have found that, if the state matrix $\boldsymbol{A}$ is symmetric strictly diagonally dominant with negative diagonal entries,  \eqref{eq:rosen} always holds. Thus, we formulate the conjecture below.

\begin{conj}[Unique Nash equilibrium]\label{conj:strict-mono}
If $\boldsymbol{A}$ is symmetric strictly diagonally dominant with negative diagonal entries, then the $n$-player noncooperative game has a unique Nash equilibrium. 
\end{conj}

We prove the conjecture for the $2$-player case. For simplicity, we assume $\rho_1=\rho_2=0$.

\begin{Proof}{\bf 2-players}
In this case, $\boldsymbol{K}-\boldsymbol{A}$ can be parameterized as
\begin{align}\label{eq:K-A-2}
   \boldsymbol{K}-\boldsymbol{A}=\begin{bmatrix}k_1-a_{11}&-a_{12}\\-a_{12}&k_2-a_{22}\end{bmatrix}  \quad \text{and}\quad \left(\boldsymbol{K}-\boldsymbol{A}\right)^{-1}=\dfrac{1}{\nu}\begin{bmatrix}k_2-a_{22}&a_{12}\\a_{12}&k_1-a_{11}\end{bmatrix},
\end{align}
with $a_{11}<-|a_{12}|$, $a_{22}<-|a_{12}|$, $k_1\in\left[0, \overline{k}_1\right]$, $k_2\in\left[0, \overline{k}_2\right]$, and $\nu:=\left(k_1-a_{11}\right)\left(k_2-a_{22}\right)-a_{12}^2$. Note that this parameterization ensures that
\begin{align}\label{eq:k-inq}
k_1-a_{11}>|a_{12}|\geq0\qquad\text{and}\qquad k_2-a_{22}>|a_{12}|\geq0 \,.
\end{align}
We next write the pseudogradient $\boldsymbol{g}(\boldsymbol{k})$ defined in \eqref{eq:g}. Observe from \eqref{eq:dJi} and \eqref{eq:dfi} in the proof of Lemma~\ref{lem:convex} that, when $\rho_i=0$, we have $\partial J_i(k_i,k_{-i})/\partial k_i=-f^2_i(k_i,k_{-i})/2$, $\forall i \in \left[n\right]$, which together with the definition of $f_i(k_i,k_{-i})$ in \eqref{eq:Ji} and the expression of $\left(\boldsymbol{K}-\boldsymbol{A}\right)^{-1}$ in \eqref{eq:K-A-2} yields
\begin{align}
\boldsymbol{g}(\boldsymbol{k})= -\dfrac{1}{2\nu^2} \begin{bmatrix}\left(k_2-a_{22}\right)^2\\\left(k_1-a_{11}\right)^2\end{bmatrix} \,.
\end{align}
Through standard calculus, we can get
\begin{align*}
    \boldsymbol{G}(\boldsymbol{k})+\boldsymbol{G}^T(\boldsymbol{k})=\dfrac{1}{\nu^3}\begin{bmatrix}2\left(k_2-a_{22}\right)^3&a_{12}^2\left(k_1-a_{11}+k_2-a_{22}\right)\\a_{12}^2\left(k_1-a_{11}+k_2-a_{22}\right)&2\left(k_1-a_{11}\right)^3\end{bmatrix}\,.
\end{align*}
By Sylvester’s criterion, to show $\boldsymbol{G}(\boldsymbol{k})+\boldsymbol{G}^T(\boldsymbol{k})\succ0$, it suffices to show
\begin{align}\label{eq:mu}
\mu:=4\left(k_1-a_{11}\right)^3\left(k_2-a_{22}\right)^3-a_{12}^4\left(k_1-a_{11}+k_2-a_{22}\right)^2>0 \,,
\end{align}
since $\nu>0$ and $k_2-a_{22}>0$ due to \eqref{eq:k-inq}. To see why \eqref{eq:mu} holds, we can expand $\mu$ as
\begin{align*}
\mu=&\ 4\left(k_1-a_{11}\right)^3\left(k_2-a_{22}\right)^3-a_{12}^4\left(k_1-a_{11}\right)^2-a_{12}^4\left(k_2-a_{22}\right)^2-2a_{12}^4\left(k_1-a_{11}\right)\left(k_2-a_{22}\right)\\=&\left(k_1-a_{11}\right)^2\left[\left(k_1-a_{11}\right)\left(k_2-a_{22}\right)^3-a_{12}^4\right]+\left(k_2-a_{22}\right)^2\left[\left(k_1-a_{11}\right)^3\left(k_2-a_{22}\right)-a_{12}^4\right]\\&+2\left(k_1-a_{11}\right)\left(k_2-a_{22}\right)\left[\left(k_1-a_{11}\right)^2\left(k_2-a_{22}\right)^2-a_{12}^4\right]>0 \,,   
\end{align*}
where the inequality is due to \eqref{eq:k-inq}. 
\end{Proof}

The proof of Conjecture~\ref{conj:strict-mono} for the general $n$-player game is a direction of future research.

%% file: 04-learning.tex

In this section, we present a mechanism for the players to reach the Nash equilibrium of the game. Since the goal of each player is to selfishly minimize its own cost $J_i(k_i,k_{-i})$ in \eqref{eq:cost-def}, an intuitive choice for each player as the game proceeds is to update its action $k_i$ by modifying it in the direction where the cost $J_i(k_i,k_{-i})$ descends the fastest. More specifically, after a random initialization of the action $k_i$, denoted as $k_i^{(0)}$, such that $k_i^{(0)}\in\left[0, \overline{k}_i\right]$, each player updates its action $k_i^{(l)}$ at the $l$th stage of the game along a projected direction of cost descent, i.e.,
\vspace{-0.2cm}
\begin{align}\label{eq:kdyn}
k_i^{(l)}= \left[k_i^{(l-1)} - \dfrac{\partial J_i(k_i,k_{-i})}{\partial k_i}\Biggr|_{(k_i^{(l-1)},k_{-i}^{(l-1)})} \right]_{0}^{\overline{k}_i}, \qquad\forall l=1,2,\dots\,,
\end{align}
where the projection $[\cdot]_a^b:=\min(\max(\cdot,a),b)\in\left[a, b\right]$ ensures that $k_i^{(l)}\in\left[0, \overline{k}_i\right]$, $\forall l=1,2,\dots$. If the Nash equilibrium is unique, then the gradient update converges to it~\citep{rosen1965Eco}. Using the results in~\citep{Ratliff13}, it is not hard to show that each of Nash equilibria are locally stable. Hence if $\boldsymbol{k}$ is initialized close to an equilibrium, it would converge to it (we skip the details here because of length constraints). 
%
%
%
%

\subsection{Implementation of the Action Updating Rule}

To implement the action updating rule~\eqref{eq:kdyn}, each player has to compute the partial derivative of its cost function $J_i(k_i,k_{-i})$ with respect to its action $k_i$ at the current stage. In general, this marginal cost is not explicitly available. Here, we describe our approach to tackle this. The following result provides an explicit expression of the partial derivative in terms of the cost itself.

\begin{prop}[Estimation of marginal cost]\label{prop:mc}
The marginal cost function of the $i$th player is
\begin{align}\label{eq:mc}
 \dfrac{\partial J_i(k_i,k_{-i})}{\partial k_i}= \dfrac{2J_i(k_i,k_{-i})}{1+\rho_i k_i^2}\left(\rho_i k_i-J_i(k_i,k_{-i})\right)  \,,\quad \forall i \in \left[n\right]\,.
\end{align}
\end{prop}
\begin{Proof}
We use the expressions of the partial derivatives of $J_i(k_i,k_{-i})$ and $f_i(k_i,k_{-i})$ with respect to $k_i$ in \eqref{eq:dJi} and \eqref{eq:dfi}, respectively, to express the marginal cost in terms of $k_i$ and $f_i(k_i,k_{-i})$. Substituting \eqref{eq:dfi} into \eqref{eq:dJi} yields
\begin{align}\label{eq:dJi-fi}
    \dfrac{\partial J_i(k_i,k_{-i})}{\partial k_i}=\rho_i k_i f_i(k_i,k_{-i})-\dfrac{\left(1+\rho_i k_i^2\right)}{2}f^2_i(k_i,k_{-i})\,.
\end{align}
Now, by Lemma~\ref{lem:Ji-ki}, we have
\begin{align}\label{eq:fi-Ji}
    f_i(k_i,k_{-i})=\dfrac{2J_i(k_i,k_{-i})}{1+\rho_i k_i^2}\,.
\end{align}
Substituting \eqref{eq:fi-Ji} into \eqref{eq:dJi-fi}, we get
\begin{align*}
    \dfrac{\partial J_i(k_i,k_{-i})}{\partial k_i}\!=\rho_i k_i \dfrac{2J_i(k_i,k_{-i})}{1+\rho_i k_i^2}-\dfrac{1+\rho_i k_i^2}{2}\!\left(\dfrac{2J_i(k_i,k_{-i})}{1+\rho_i k_i^2}\right)^2\!\!=\dfrac{2J_i(k_i,k_{-i})}{1+\rho_i k_i^2}\!\left(\rho_i k_i-J_i(k_i,k_{-i})\right)\!\,,
\end{align*}
concluding the proof.
\end{Proof}

Proposition~\ref{prop:mc} indicates that the marginal cost for a given action $k_i$ is computable through \eqref{eq:mc} as long as each player knows the values of its own cost $J_i(k_i,k_{-i})$ for that action at the current stage.
%
%
A way to do this becomes clear when recalling the original definition of $J_i(k_i,k_{-i})$ in \eqref{eq:cost-def} as a selfish expected cost-to-go on state deviations and control efforts over an infinite time-horizon. It readily follows that each player can estimate $J_i(k_i,k_{-i})$ by averaging its own cost-to-go along a batch of sampled trajectories over a finite time-horizon. Specifically, $\forall l=1,2,\dots$, at the $l$th stage of the game, each player can estimate $J_i(k_i^{(l-1)},k_{-i}^{(l-1)})$ via 
\begin{align}\label{eq:cost-est}
J_i(k_i^{(l-1)},k_{-i}^{(l-1)})\approx&\ \dfrac{1}{|\mathcal{B}|}\sum_{b=1}^{|\mathcal{B}|}\! \left[\int_{0}^{T_\mathrm{s}}\!\!\left(x_i^2(t)+\rho_i u_i^2(t)\right)\!\!\ \mathrm{d}t\right]^{\langle b \rangle}\Biggr|_{(k_i^{(l-1)},k_{-i}^{(l-1)})} \,,\quad\forall i \in \left[n\right],
\end{align}
where $|\mathcal{B}|$ is the batch size, $T_\mathrm{s}$ is the sampling time-horizon, and, with an abuse of notation, we simply introduce a superscript $\langle b \rangle$ to denote the cost along the $b$th trajectory in the batch rather than accurately distinguishing $x_i(t)$ and $u_i(t)$ along different trajectories to avoid complicating the notation. As $|\mathcal{B}|\to\infty$, the error in using \eqref{eq:cost-est} goes to zero (law of large numbers). The rate can be bounded if we assume more information on the distribution of $\boldsymbol{x}(0)$. For example, if it has bounded moments, then the error in gradient estimate goes to zero exponentially fast~\citep{van2000asymptotic}. 

Note that all trajectories in \eqref{eq:cost-est} are generated by the system \eqref{eq:sys} given that control input from each player is $u_i(t)=-k_i^{(l-1)}x_i(t)$, with the initial condition $\boldsymbol{x}(0)$ being randomly drawn from $n$ uniform i.i.d. on $(-\sqrt{12}/2, \sqrt{12}/2)$ that have $0$ as mean and $1$ as variance such that the assumption $\mathbb{E} \left[\boldsymbol{x}(0) \boldsymbol{x}(0)^T\right]=\boldsymbol{I}_n$ holds.
%
%
Once $J_i(k_i^{(l-1)},k_{-i}^{(l-1)})$ has been estimated by \eqref{eq:cost-est}, the marginal cost can be calculated through \eqref{eq:mc} as
\begin{align}\label{eq:mc-l}
 \dfrac{\partial J_i(k_i,k_{-i})}{\partial k_i}\Biggr|_{(k_i^{(l-1)},k_{-i}^{(l-1)})}= \dfrac{2J_i(k_i^{(l-1)},k_{-i}^{(l-1)})}{1+\rho_i \left(k_i^{(l-1)}\right)^2}\left(\rho_i k_i^{(l-1)}-J_i(k_i^{(l-1)},k_{-i}^{(l-1)})\right) .
\end{align}
Equipped with this, each player can update its action $k_i^{(l)}$ at the $l$th stage via \eqref{eq:kdyn}.

%
%

\begin{rem}[Decentralized action update]
{\rm
The implementation of the action updating rule \eqref{eq:kdyn} through \eqref{eq:cost-est} and \eqref{eq:mc-l} is a repeated procedure that includes two phases in each stage, where in the first phase each player collects its own trajectories to evaluate \eqref{eq:cost-est} and \eqref{eq:mc-l} for the given actions, and in the second phase all players execute \eqref{eq:kdyn}. This procedure is decentralized in the sense that each player only needs to estimate its own marginal cost by observing its own sampled trajectories, without knowing the actions of its
opponents, although the evolution of the trajectories
depends on the actions taken by all players.
} \hfill $\square$
\end{rem}
%
%

\subsection{Experiments of the Action Updating Rule in Noncooperative Game}

With the implementation of the action updating rule \eqref{eq:kdyn} being made explicit, we test its performance in the noncooperative game. For a test involving $5$ players, we randomly generate a symmetric strictly diagonally dominant matrix $\boldsymbol{A}\in \real^{5\times 5}$ with negative diagonal entries as
\begin{align*}
\boldsymbol{A}=\begin{bmatrix}-0.0342&-0.0111 & \ \ \ 0.0095 & -0.0012 &  \ \ \ 0.0118\\-0.0111 & -0.0627 &  \ \ \ 0.0098 &  \ \ \ 0.0155 &  \ \ \ 0.0254\\ \ \ \ 0.0095 &  \ \ \ 0.0098 & -0.0341 & -0.0065 & -0.0081\\-0.0012 &  \ \ \ 0.0155 & -0.0065 & -0.0323 & -0.0081\\ \ \ \ 0.0118 &  \ \ \ 0.0254 & -0.0081 
&-0.0081 & -0.1086 
\end{bmatrix}.
\end{align*}
The tradeoff coefficient $\rho_i$ of each player is also generated randomly from $(0,1)$ as: $\rho_1=0.5542$, $\rho_2=0.2642$, $\rho_3=0.4526$, $\rho_4=0.0664$, and $\rho_5=0.7990$. As the game kicks off, each player randomly initializes a positive action $k_i^{(0)}$ and updates its action $k_i^{(l)}$, $\forall l=1,2,\dots$, based on \eqref{eq:kdyn} with batch size $|\mathcal{B}|=500$ and sampling time-horizon $T_\mathrm{s}=\SI{200}{\second}$. The upper bound $\overline{k}_i$ on action is set to be sufficiently large such that it is never activated.
Fig.~\ref{fig:k-traj} plots the evolution of individual actions, costs, and gradients in two different rounds of the aforementioned game, which shows that, although each player initializes its action differently, the same equilibrium is always reached where individual gradients all converge to zero. Table~\ref{table:k} confirms this observation by showing numerically that the final actions $k_i^{(250)}$ of each player are practically the same while their initial values $k_i^{(0)}$ are different in each round.


\begin{figure}[!t]
\centering
\includegraphics[width=0.325\columnwidth]{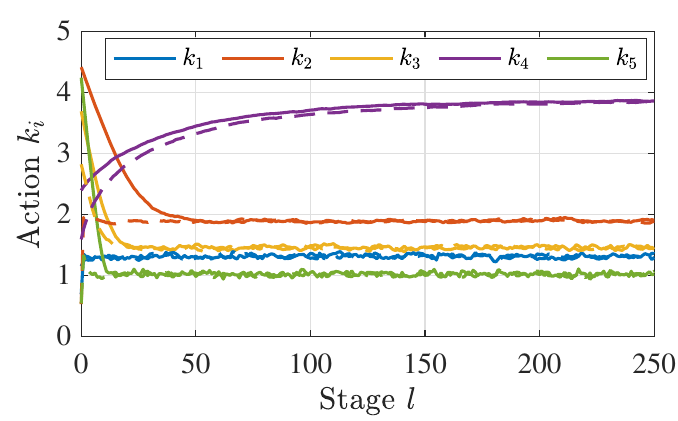}\label{fig:k-traj}
\includegraphics[width=0.325\columnwidth]{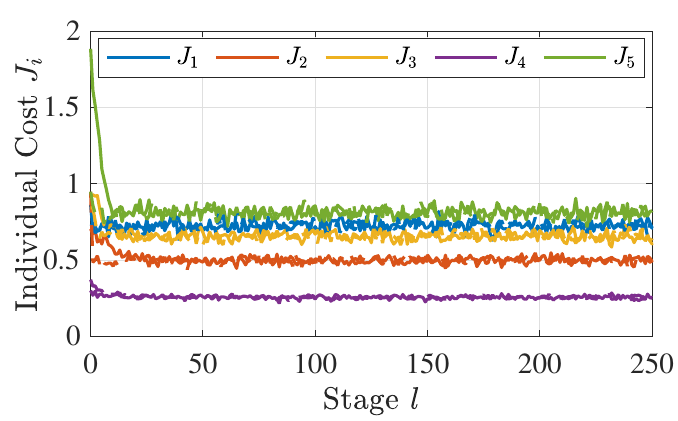}\label{fig:diag3}
\includegraphics[width=0.325\columnwidth]{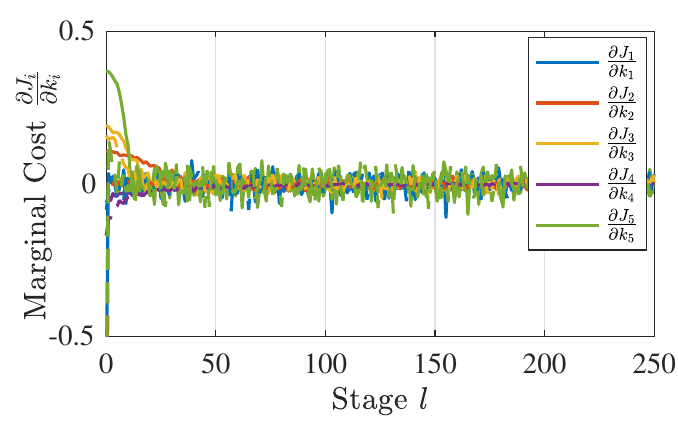}\label{fig:diag3}
\caption{Evolution of individual actions (left), costs (center), and gradients (right) under the updating rule~\eqref{eq:kdyn} with different initializations. The trajectories for different initializations are distinguished by solid and dashed lines.}
\label{fig:k-traj}
\end{figure}

\begin{table}[!t]
\renewcommand{\arraystretch}{1.2}
\centering
\caption{Comparison Between Two Rounds of the Game}
\begin{threeparttable}[t]
\begin{tabular}{c|c|c|c|c|c|c}
\hline\hline
\multicolumn{2}{c|}{\multirow{2}{*}{Action}}& \multicolumn5{c}{Player}\\\cline{3-7}\multicolumn{2}{c|}{}&$1$&$2$&$3$&$4$&$5$\\
\hline
\multirow{2}{*}{$k_i^{(0)}$}&Round $1$&$0.69$ &$4.41$ &$3.69$&$2.39$&$4.24$\\\cline{2-7}
&Round $2$ &$1.15$ &$0.53$ &$2.82$&$1.59$&$0.54$\\
\hline
\multirow{2}{*}{ $k_i^{(250)}$}&Round $1$&$1.31$ &$1.89$ &$1.46$&$3.85$&$1.03$\\\cline{2-7}
&Round $2$ &$1.29$ &$1.88$&$1.49$&$3.85$&$1.03$\\
\hline\hline
\end{tabular}
\end{threeparttable}
\label{table:k}
\end{table}